# Beyond the Small-Angle Approximation For MBR Anisotropy from Seeds[1]

ALBERT STEBBINS♠ and SHOBA VEERARAGHAVAN♣

♠ NASA/Fermilab Astrophysics Center

*stebbins@perseus.fnal.gov*

♣ Goddard Space Flight Center

*shoba@clb.gsfc.nasa.gov*

ABSTRACT. In this paper we give a general expression for the energy shift of massless particles travelling through the gravitational field of an arbitrary matter distribution as calculated in the weak field limit in an asymptotically flat space-time. It is *not* assumed that matter is non-relativistic. We demonstrate the surprising result that if the matter is illuminated by a uniform brightness background that the brightness pattern observed at a given point in space-time (modulo a term dependent on the oberver's velocity) depends only on the matter distribution on the observer's past light-cone. These results apply directly to the cosmological MBR anisotropy pattern generated in the immediate vicinity of of an object like a cosmic string or global texture. We apply these results to cosmic strings, finding a correction to previously published results for in the small-angle approximation. We also derive the full-sky anisotropy pattern of a collapsing texture knot.

## I. Introduction

Recent measurements of anisotropies in the Microwave Background Radiation (MBR) (see refs [1]) are clearly a great advance in our understanding of the universe around us. MBR anisotropies may provide us with our only direct probe of the structure of the universe on the largest scales accessible by observation. Unfortunately in trying to obtain an unambiguous interpretation of MBR anisotropies one is beset by myriad uncertainties. Microwave emission from Galactic and extra-Galactic source make the measurement of the primordial anisotropies difficult, especially on small angular scales. In the finite part of our universe which is observable we will inevitably have problems with finite sampling (a.k.a. cosmic variance) in determining the statistical properties of inhomogeneities on large scales. Uncertainties in ionization history and the fundamental cosmological parameter (e.g. $\Omega_0$ and $\Lambda_0$) give large uncertainties in relating the anisotropies to the density inhomogeneities. Compounding these uncertainties is the unknown nature of the cosmological inhomogeneities. These may be primordial adiabatic or isocurvature, or may be non-primordial seeded perturbations. The primordial density perturbations may be accompanied by gravitational waves or even vorticity. The statistics may be Gaussian or not, and the spectrum of perturbations may take various different forms. Unambiguously determining the meaning of the MBR anisotropies will be a difficult task!

Given these uncertainties it is interesting to note that in certain classes of seeded perturbations very distinctive signatures in the pattern of MBR anisotropy will be left by the seeds

---

themselves. For example cosmic strings will lead to temperature discontinuities along curves on the sky (refs [2]) while cosmic textures can lead to both hot and cold spots in the temperature pattern (refs [3]). The reason that such distinctive features are produced is that the perturbations are not primordial but are seeded at recent times. These distinctive features are the imprints of the seeds "in the act" of seeding the inhomogeneities. If definitive evidence for the presence of such distinctive features were found this would provide strong evidence for the presence of such seeds.

In this paper we continue the study of the nature of the anisotropies produced by seeds. In particular we will derive an explicit expressions for the all-sky anisotropy for an arbitrary seed distribution history and apply this to a few specific seed configurations. However we do not do this in a cosmological setting but rather our calculation is for an isolated source in a Minkowski background illuminated by a constant temperature background of photons. This idealization is useful and interesting because it leads to a very simple expression for which analytic expressions are easy to come by. An extension of this result to an expanding universe will be given in a subsequent paper (ref [4]) where the expressions are much more complicated. The results provided here can be directly applied to seeds at low redshift ($z \lesssim 1$) where the cosmological effects can be ignored. The small-angle limit of the results also applies for light-rays which pass a seed with impact parameters much smaller than the horizon at the time of passing. Thus, for example, the temperature discontinuity across a string was derived in Minkowski space but applies equally well to a cosmological string, since the discontinuity is a small-angle effect. What is not so obvious is that the small-angle limit of the Minkowski formula, without any restrictions of the angle being much smaller than the horizon, applies nearly exactly to the cosmological case. The only part which is missing from the small-angle Minkowski formula is the term due to the time changing potential induced by the decaying modes set up by the seed. We will show this in ref [4]. Thus the small-scale anisotropies in Minkowski pace should be indicative of what we will find in an expanding universe. Of course there can be no exact correspondence since the seeds will evolve differently in a cosmological setting.

A proper analysis of small-scale cosmological anisotropies must include the treatment of multiple scattering of the photons which has been worked out by many authors (ref [5]). We shall not deal with these effects here. The physics of large-scale MBR anisotropies is given by the geodesic equation in linearized gravity and was first worked out by Sachs and Wolfe in ref [6]. The general result can be written as

$$\frac{\Delta T}{T} = \frac{\Delta T}{T}_\text{i} + \frac{1}{c^2}\left(\Phi_\text{em} - \Phi_\text{obs}\right) - \frac{1}{c}\hat{\mathbf{n}}\cdot\left(\mathbf{v}_\text{em} - \mathbf{v}_\text{obs}\right) + \frac{2}{c^2}\int_{t_\text{em}}^{t_\text{obs}} \dot{\Phi}_\text{ISW}\, dt \qquad (1.1)$$

where the unit vector, $\hat{\mathbf{n}}$, gives the direction in which one is measuring the temperature, $\Delta T/T_\text{i}$ gives the fractional deviation from the temperature anisotropy at the point of emission (or last scattering); $\Phi_\text{em}$ and $\Phi_\text{obs}$ are the gravitational potential at emission and observation, respectively; $\mathbf{v}_\text{em}$ and $\mathbf{v}_\text{obs}$ gives the velocity with respect to the cosmic rest frame of the emitter and observer, respectively; and the last term is an integral along the photon trajectory of the time rate of change of (the appropriate component of) the gravitational field. In situations with non-relativistic matter $\Phi_\text{ISW}$ is just the gravitational potential. This last term is often referred to as the "integrated Sachs-Wolfe" (ISW) effect. Eq (1.1) applies equally well in Minkowski space or a Friedmann-Robertson-Walker universe. In a matter-dominated universe with adiabatic growing mode perturbations $\Delta T/T_\text{i} = -2\Phi_\text{em}/(3c^2)$, $\dot{\Phi}_\text{ISW} = \dot{\Phi} = 0$ leading to the well known formula

$$\frac{\Delta T}{T} = +\frac{1}{3c^2}\Phi_\text{em} + \frac{1}{c}\hat{\mathbf{n}}\cdot\mathbf{v}_\text{em} + \cdots \qquad (1.2)$$

where the $\cdots$ refer to the effect of observer potential and velocity which only contribute to the monopole and dipole components of the anisotropy. The analysis here will concern asymptotically flat space where $\Delta T/T_\mathrm{i} = \Phi_\mathrm{em} = \mathbf{v}_\mathrm{em} = 0$ so we are left with

$$\frac{\Delta T}{T} = \frac{\Delta T}{T}_\mathrm{i} - \frac{1}{c^2}\Phi_\mathrm{obs} + \frac{1}{c}\hat{\mathbf{n}}\cdot\mathbf{v}_\mathrm{obs} + \frac{1}{c^2}\int_{t_\mathrm{em}}^{t_\mathrm{obs}} \dot{\Phi}_\mathrm{ISW}\, dt. \tag{1.3}$$

Thus apart from the monopole and dipole term the effect we are calculating is the ISW effect which is relatively unimportant in a more conventional scenario. However it is this term which is likely to dominant in scenarios with seeds.

The rest of the paper will proceed as follows: in

§2 we give the expression for the anisotropy pattern produced by an arbitrary seed configuration, in

§3 we discuss various geometrical properties of this result, in

§4 we take the small-angle limit, in

§5 we give the temperature pattern for a moving point mass, in

§6 we apply our formulae to cosmic strings, in

§7 formulae are given for the anisotropy averaged on circles on the sky, in

§8 these formula are applied to a collapsing texture knot, in

§9 a summary of results is given, and in the

*Appendix* a brief outline of the derivation of the main formula is given.

For those interested only in the main result §2 and §9 are recommended.

## 2. Sachs-Wolfe Integral for an Arbitrary Matter Distribution in Flat Space

Here we will calculate the change in the energy distribution of initially thermal distribution of photons due to the gravitational field generated by an arbitrary matter distribution. In particular we consider an observer in an asymptotically flat space-time which contains some distribution of matter. This matter distribution we will refer to as the *source*. Let us suppose that at some early time a thermal distribution of photons is released which has the same temperature everywhere. Furthermore we assume that there is negligible direct interactions of the photons with the matter, e g. via refraction, reflection, opacity, etc. However the photons must travel through the gravitational field of the source before they reach the observer which will change the energy of the photons and thus the energy distribution received by the observer will not be the same in all directions at all times. It is well known that a gravitational field cannot change the thermal nature of the energy distribution along any given light-ray, but will only effect the temperature characterizing this distribution, (see refs [7]). The temperature change is just related to the energy shift of any given photon in that distribution by the relation

$$\frac{\Delta T}{T} = \frac{T_\mathrm{obs} - T_\mathrm{em}}{T_\mathrm{em}} = \frac{E_\mathrm{obs} - E_\mathrm{em}}{E_\mathrm{em}} \tag{2.1}$$

where $E_{\rm em}$ and $E_{\rm obs}$ are the energy of the photon at emission and observation respectively. The observer will see different temperatures in different directions in the sky and below we will give an expression for these different temperatures as a function of the observer position, $\mathbf{x}_{\rm obs}$, the observation time, $t_{\rm obs}$, and the direction of observation, $\hat{\mathbf{n}}$. In this way we are specifying the complete photon distribution in all of space-time. Of particular interest is the temperature shift observed at a given place and time, i.e. how the temperature shift varies with $\hat{\mathbf{n}}$ for fixed $\mathbf{x}_{\rm obs}$ and $t_{\rm obs}$. This is why $\Delta T/T$ is often referred to as *anisotropy* which is what we will call it below.

### Assumptions, Notation, and Conventions

We set the speed of light, $c$, and Newton's constant, $G$ to unity in what follows. Our tensor notation uses greek letters for the 4 space-time indices, and Latin letters for the 3-spatial indices. Einstein's index summation convention is used throughout. The gravitational field $g_{\mu\nu}$ of a particular matter distribution depends only on the stress-energy distribution of that matter, which we denote by $\Theta_{\mu\nu}(\mathbf{x}, t)$. We will *not* require that the matter be non-relativistic, i.e., we *do not* require $\Theta_{ij} \ll \Theta_{i0} \ll \Theta_{00}$. However we will assume that the gravitational fields are weak, and require that the matter distribution respect this condition. In the weak-field approximation, the metric is, $g_{\mu\nu} = \text{diag}[-1, 1, 1, 1] + h_{\mu\nu}$, with $h_{\mu\nu} \ll 1$, and we need calculate the photon energy shift only to first order in $h_{\mu\nu}$. The weak field approximation lets us integrate the photon geodesic equation along the unperturbed trajectory, and to evolve the matter distribution in the vacuum (Minkowski) background. Thus the various components of $\Theta_{\mu\nu}$ have the usual meaning in a given Minkowski frame: $\Theta_{00}$ is the density, $-\Theta_{i0}$ is the energy flux or momentum density, and $\Theta_{ij}$ is the pressure (stress) tensor.

The temperature is only defined with respect to a given inertial frame. Above we have stated that at large distance and early times the distribution is thermal. By this we mean that it is an isotropic thermal distribution at rest in a given global rest frame. We may define such a rest frame in the asymptotically flat part of the space-time. Since we have restricted ourselves to weak gravitational fields this global rest frame is defined throughout the space-time up to small (non-relativistic) Lorentz boosts. The weak field also guarantees that the anisotropy is also small ($\ll 1$). Small Lorentz boosts of a nearly isotropic temperature field only changes the dipole ($l = 1$ in a spherical harmonic expansion) part of the temperature anisotropy while all other components of the temperature pattern are frame (or coordinate or gauge) independent in the weak field limit. The monopole (or mean or $l = 0$) component of the anisotropy is also coordinate independent so long as one restricts oneself to localized perturbations. Of course the observer may not know the asymptotic temperature at large distance and may therefore have no fiducial temperature with which to compare to. In this sense the monopole component may be considered to also be unobservable.

In fact one can make sense of the dipole anisotropy in a coordinate independent way if the source is moving with respect to the global rest frame defined above. In this case the space is asymptotically flat not only at large distances at a fixed time, but also at early times ($t \to -\infty$) at a fixed position. One can then uniquely define a congruence of freely falling observer which were at rest with respect to the thermal photon distribution at early times. These observers then define a unique rest frame at all times and one may calculate the temperature pattern in this rest frame. This is the "definition" of the dipole anisotropy which will use.

### General Solution

By first solving for the metric perturbation in terms of the stress-energy distribution and substituting this into the Sachs-Wolfe integral represented in eq (1.1) we obtain expression for the temerature pattern seen by freely falling observer:

$$\frac{\Delta T}{T}(\hat{\mathbf{n}}, \mathbf{x}_{\text{obs}}, t_{\text{obs}}) = \int d^3x' \left[ 2 \frac{X_{\text{obs}} \hat{n}^i + X_{\text{obs}}^i}{X_{\text{obs}}(X_{\text{obs}} + \hat{\mathbf{n}} \cdot \mathbf{X}_{\text{obs}})} \left( \Theta_{0i}(\mathbf{x}', t_{\text{obs}} - X_{\text{obs}}) - \hat{n}^j \Theta_{ij}(\mathbf{x}', t_{\text{obs}} - X_{\text{obs}}) \right) \right.$$

$$+ \frac{X_{\text{obs}} - \hat{\mathbf{n}} \cdot \mathbf{X}_{\text{obs}}}{X_{\text{obs}}^2} \Theta_+(\mathbf{x}', t_{\text{obs}} - X_{\text{obs}}) - 4 \frac{1}{X_{\text{obs}}} \hat{n}^i \Theta_{0i}(\mathbf{x}', t_{\text{obs}} - X_{\text{obs}})$$

$$\left. - \frac{\hat{\mathbf{n}} \cdot \mathbf{X}_{\text{obs}}}{X_{\text{obs}}^3} \int\limits_{-\infty}^{t_{\text{obs}} - X_{\text{obs}}} dt' \, \Theta_+(\mathbf{x}', t') \right] \tag{2.2}$$

where $\mathbf{X}_{\text{obs}} = \mathbf{x}_{\text{obs}} - \mathbf{x}'$ is the distance vector between the observer and the source at $\mathbf{x}'$ with absolute value $X_{\text{obs}} = |\mathbf{X}_{\text{obs}}|$. An outline of the derivation of this is given in the appendix. We have used certain assumptions to derive eq (2.2), in particular

$$\lim_{t \to -\infty} \Theta_{\mu\nu}(\mathbf{x}_{\text{obs}} - \hat{\mathbf{n}}t, t) = 0 \qquad \vee \, \hat{\mathbf{n}} \qquad |\hat{\mathbf{n}}| = 1 \tag{2.3}$$

which guarantees asymptotic flatness at large distance which we require, and

$$\lim_{t \to -\infty} \Theta_{\mu\nu}(\mathbf{x}, t) = 0. \tag{2.4}$$

which guarantees the asymptotic flatness at early times. One will satisfy eq (2.3) as long as all of the source distribution moves at speeds less than the speed of light, and one will satisfy eq (2.4) if all of the source have a non-zero velocity. The first condition, eq (2.3), is really necessary to obtain any well-defined anisotropy pattern, but the second condition is only required to make sense of the dipole component of the anisotropy. If eq (2.4) were not satisfied, the integral in the last term in eq (2.2) might diverge, but we see that this integral only contributes to the dipole. If eq (2.4) is satisfied we may take $\lim_{t \to -\infty} h_{ij} = 0$ and since we are using comoving coordinates the dipole given by eq (2.2) is that which would be observed by the freely falling observer described above.

Notice the profound simplicity of the Sachs-Wolfe formula: except for the last term which only contributes to the dipole; only sources *on* the past light cone of the observer can contribute to the observable temperature distortion. This is rather unexpected in that the source configuration on the past light cone could have been produced by any one of an infinite number of different stress-energy histories, yet the exact source history is not important. Causality would allow a dependence on the source in the interior of the past light cone since the photons must pass through the gravitational field produced inside the light cone. However we find that when one sums the anisotropy produced by the gravitational field produced by the stress-energy inside the past lightcone that the sum yields only a dipole anisotropy pattern. This surprising result was found in the small-angle approximation in in ref [9]. A simple (mathematical) reason for this reduced dependence in the small angle limit has been found by Hindmarsh in ref [10] although so far we know of no generalization of this argument for the large angle case. As will be shown in ref [4] the lack of dependence on the interior of the light-cone will not extend to the expanding universe case.

The third term in eq (2.2) is a sum of a pure monopole (independent of $\hat{\mathbf{n}}$) and pure dipole (proportional to $\hat{\mathbf{n}}$) term while the last two terms in eq (2.2) are pure dipole terms. As discussed above the monopole and dipole term are not very interesting. The dipole depends on the observer's velocity and the monopole in addition to not contributing to an *anisotropy* also may be solely a measure of the local gravitational potential (see §3). The more interesting quadrupole and higher order anisotropy are contained in the remaining term

$$\frac{\Delta T}{T}(\hat{\mathbf{n}}, \mathbf{x}_{\text{obs}}, t_{\text{obs}}) = 2 \int d^3 x' \frac{X_{\text{obs}} \hat{n}^i + X^i_{\text{obs}}}{X_{\text{obs}}(X_{\text{obs}} + \hat{\mathbf{n}} \cdot \mathbf{X}_{\text{obs}})} \left[ \Theta_{0i}(\mathbf{x}', t_{\text{obs}} - X_{\text{obs}}) - \hat{n}^j \Theta_{ji}(\mathbf{x}', t_{\text{obs}} - X_{\text{obs}}) \right].$$
(2.5)

This expression does in fact still contain some residual dipole anisotropy which may be subtracted explicitly if needed.

Eq (2.2) is is the main result of this paper. While it does assume that the space-time is asymptotically flat and that the gravitational fields are weak, there are no further assumptions. In particular, we do not assume either that the sources are very far from the observer, that the angle between the lines of sight to the sources and photons are small (the small-angle approximation), or that the matter distribution is non-relativistic.

## 3. Geometrical Decomposition of the Anisotropy

The Green functions given in §2 give us the temperature pattern which is generated by each infinitesimal element of the source stress-energy distribution. Thus for each component of the stress-energy tensor at each point in space-time it gives us the incremental temperature anisotropy as a function of position on the sky which is added by that part of the stress-energy. We will now show that from simple geometrical considerations that the angular dependence must have a fairly simple form. This incremental anisotropy is a scalar function and therefore the functional dependence on the direction vector, $\hat{\mathbf{n}}$, can only be via this vector contracted with some other tensor. The only tensors that can appear in the problem are $\mathbf{X}_{\text{obs}}$, the displacement of the observer from the source point, and the various components of the stress-energy tensor. Thus the incremental anisotropy can only depend on $\hat{\mathbf{n}}$ through the combinations

$$\hat{\mathbf{n}} \cdot \mathbf{X}_{\text{obs}} \qquad \hat{n}^i \Theta_{0i} \qquad \hat{n}^i X^j_{\text{obs}} \Theta_{ij} \qquad \hat{n}^i \hat{n}^j \Theta_{ij} \qquad (3.1)$$

and we have left out the trivial $\hat{\mathbf{n}} \cdot \hat{\mathbf{n}} = 1$. Now since we are doing the calculation in linear theory the incremental anisotropy can only depend linearly on the last three combination. Thus while there is no restriction on dependence on the angle between $\hat{\mathbf{n}}$ and $\mathbf{X}_{\text{obs}}$ there is a severe restriction on the dependence on the azimuthal angle around the $\mathbf{X}_{\text{obs}}$ direction, i.e. it can written as a finite Fourier series of terms, $e^{im\phi}$, with $m \leq 2$. Thus the Green functions have a rather simple geometrical form. It is easy to see that this simple form applies equally well to any isotropic background metric and in particular to open, closed, or flat cosmological models. It is curious to note that in the small angle approximation only terms with $m \leq 1$ appear (see §4 or refs [9,10]) while in general $m = \pm 2$ might have. A more detailed study of the geometrical properties of the Green functions will be given in ref [11], where it will be shown how one make take advantage of the simplicity in numerical computations of anisotropy patterns. While the Green functions and thus the incremental anisotropy have a simple form this does not lead to any restrictions on the total anisotropy pattern. *Any* temperature pattern will be produced by *some* source stress-energy distribution.

## Monopole and Dipole and the Newtonian Limit

As argued in §2, in the calculation we are doing the monopole and dipole terms do have physical meaning. For example in the Newtonian limit we would interpret the monopole anisotropy as a measure of (minus) the Newtonian potential at the position of the observer (taking the potential at infinity to be zero) and the dipole as a measure of observer's velocity with respect to the global rest-frame of the photons. Even though we are not working in the Newtonian limit we may define and effective gravitational potential and velocity by the monopole and dipole, i.e.

$$\frac{\Delta T}{T}(\hat{\mathbf{n}}, \mathbf{x}_{\text{obs}}) = -\Phi_{\text{eff}} + \hat{\mathbf{n}} \cdot \mathbf{v}_{\text{eff}} + \text{higher order terms} \qquad (3.2)$$

where

$$\Phi_{\text{eff}} = -\int d^3\mathbf{x}' \frac{1}{X_{\text{obs}}} \left[ \Theta_+ + 2\frac{X^i_{\text{obs}}}{X_{\text{obs}}}\Theta_{oi} + \frac{X^i_{\text{obs}} X^j_{\text{obs}}}{X^2_{\text{obs}}} \Theta_{ij} \right]$$

$$v^i_{\text{eff}} = -\int d^3\mathbf{x}' \frac{1}{X_{\text{obs}}} \left[ \frac{X^i_{\text{obs}}}{X_{\text{obs}}} \left( \Theta_+ + 3\frac{X^j_{\text{obs}}}{X_{\text{obs}}}\Theta_{oj} + 2\frac{X^j_{\text{obs}} X^k_{\text{obs}}}{X^2_{\text{obs}}} \Theta_{jk} + \frac{1}{X_{\text{obs}}} \int_{-\infty}^{t_{\text{obs}} - X_{\text{obs}}} dt' \, \Theta_+(\mathbf{x}', t') \right) \right.$$

$$\left. + \left( \Theta_{oi} + \frac{X^j_{\text{obs}}}{X_{\text{obs}}} \Theta_{ij} \right) \right]$$

(3.3)

The integral in $\mathbf{v}_{\text{eff}}$ might diverge if one does not enforce eq (2.4), for just the reasons discussed in §2. For a moving source this is not a problem. One might find it curious that the leading contribution to $\mathbf{v}_{\text{eff}}$ at large distances goes like $1/X_{\text{obs}}$ while the Newtonian gravitational acceleration goes as $1/X^2_{\text{obs}}$. One should note however that for a moving source the time integral of the gravitational acceleration really does go like $1/X_{\text{obs}}$ since the relevant timescale over which the most of the acceleration takes place is proportional to the distance, i.e. $\Delta t \sim X_{\text{obs}}/v$. Thus this is not really a different scaling than in Newtonian gravity.

The limit in the case of non-relativistic sources, i.e. $\Theta_{ij} \ll \Theta_{oi} \ll \Theta_{00}$, we find

$$\Phi_{\text{eff}} \approx -\int d^3\mathbf{x}' \frac{1}{X_{\text{obs}}} \Theta_{00} = \Phi_{\text{Newt}}$$

$$\mathbf{v}_{\text{eff}} \approx -\int d^3\mathbf{x}' \frac{\mathbf{X}_{\text{obs}}}{X^2_{\text{obs}}} \left[ \Theta_{00} + \frac{1}{X_{\text{obs}}} \int_{-\infty}^{t_{\text{obs}} - X_{\text{obs}}} dt' \, \Theta_{00}(\mathbf{x}', t') \right]$$

(3.4)

which is not quite the Newtonian result. However for a non-relativistic source we must also require that the velocity of the source be small, i.e. $v \ll 1$, in which case the time interval over which the integral contributes is long enough that the integral term dominates, i.e.

$$\mathbf{v}_{\text{eff}} \approx -\int d^3\mathbf{x}' \frac{\mathbf{X}_{\text{obs}}}{X^3_{\text{obs}}} \int_{-\infty}^{t_{\text{obs}} - X_{\text{obs}}} dt' \, \Theta_{00}(\mathbf{x}', t') = -\int_{-\infty}^{t_{\text{obs}} - X_{\text{obs}}} dt' \, \nabla\Phi = \mathbf{v}_{\text{Newt}} \qquad (3.5)$$

which is the Newtonian result, i.e. that the velocity is just the time integral of the Newtonian acceleration. Thus our relativistic calculation recovers the Newtonian result for non-relativistic sources.

It is interesting that the incremental contribution to $\mathbf{v}_{\text{eff}}$ is not always directed directly toward the source point, i.e. $\mathbf{v}_{\text{eff}}$ is not parallel to $\mathbf{X}_{\text{obs}}$. In addition to attracting the observer toward the source (or possibly repelling from if the weak energy condition if violated) there is an effective "frame dragging". The last term in eq (3.5) gives a contribution to $\mathbf{v}_{\text{eff}}$ which is the direction of $-\Theta_{0i} - X_{\text{obs}}^j \Theta_{ij}/X_{\text{obs}}$ which is approximately the direction of the momentum density. Crudely speaking this is the same sign as one might expect from Mach's principle. Note that it is the opposite sign from what one might expect from the extrapolating small angle results. In the small angle approximation a moving object yields a negative temperature decrement in front of it and a positive temperature increment in back of it (see refs [2,9]) which one might think would tend to contribute to $\mathbf{v}_{\text{eff}}$ in the direction opposite to the momentum density of the source. The large angle structure of the Green function of eq (2.2) invalidates this extrapolation.

## 4. The Small-Angle Limit

Of particular interest is the small-angle limit of the Green functions of §2. For most cosmological MBR anisotropy experiments the differences in temperature are really only measured in a very small region of the sky where the small-angle formulae should give a good approximation. In §2 we have calculated the energy shift along an arbitrary light-like geodesic. Of course a single geodesic cannot be considered either small-angle or large-angle. The small-angle approximation is a reference to which geodesics one is comparing when one is calculating the anisotropy, i.e. the temperature difference. In addition to meaning that the geodesics are close to each other, the small-angle approximation usually also means that the geodesics are parallel to each other. Thus what one calculates is the energy shift on a plane of photons moving perpendicular to the plane. The temperature pattern one obtains is that which one would see if this plane of photons were projected onto a screen. How this differs from the anisotropies considered in §2-3 is that the photons which are being compared do not converge to a point at the observer. This version of the small-angle approximation is what is used in ref [10]. In ref [9] the additional assumption is made that the observer is at large distance from the source essentially in the asymptotically flat part of the space-time.

To begin we will assume that the monopole and dipole have been explicitly subtracted from the anisotropy field and thus use eq (2.5) rather than eq (2.2). These terms would contribute negligibly to temperature differences between nearby directions in any case. If we assume that the angle between the direction to the source, $-\mathbf{X}_{\text{obs}}$, and the direction from which the photon is coming, $\hat{\mathbf{n}}$, is small then

$$\frac{X_{\text{obs}}\hat{\mathbf{n}} + \mathbf{X}_{\text{obs}}}{X_{\text{obs}}(X_{\text{obs}} + \hat{\mathbf{n}}\cdot\mathbf{X}_{\text{obs}})} \to 2\frac{\mathbf{X}_{\text{obs}}^\perp}{X_{\text{obs}}^{\perp\,2}} \qquad \mathbf{X}_{\text{obs}}^\perp \equiv \mathbf{X}_{\text{obs}} - (\hat{\mathbf{n}}\cdot\mathbf{X}_{\text{obs}})\hat{\mathbf{n}} \qquad (4.1)$$

so that

$$\frac{\Delta T}{T} \approx -4\int d^3x' \frac{\mathbf{X}_{\text{obs}}^\perp}{X_{\text{obs}}^{\perp\,2}} \cdot \mathbf{U} \qquad U^i(\mathbf{x}') = -\Theta_{0i}(\mathbf{x}', t_{\text{obs}} - X_{\text{obs}}) + \hat{n}_0^j \Theta_{ij}(\mathbf{x}', t_{\text{obs}} - X_{\text{obs}}). \qquad (4.2)$$

Only $m = \pm 1$ terms are present in the small angle approximation while the $m = 0$ and $m = \pm 2$ corrections are important only for larger angles.

Note that eq (4.2) gives the anisotropy in the same form as in ref [9]. As noted in §VIg of ref [9] one may use 2-dimensional potential theory to rewrite eq (4.2) in a particularly simple form

$$\nabla_\perp^2 \frac{\Delta T}{T} \approx -8\pi \nabla_\perp \cdot \int_0^{t_{\rm obs}} dr\, {\bf U}({\bf x}_{\rm obs} + {\bf X}_{\rm obs}^\perp + \hat{\bf n}_0 r), \qquad \nabla_\perp = \nabla - \hat{\bf n}(\hat{\bf n}\cdot\nabla) \qquad (4.3)$$

which for an isolated source is equivalent to eq (4.2) when combined with boundary condition that the anisotropy go to zero for large $X_{\rm obs}^\perp$. Eq (4.3) is particularly useful since it shows a simple way of numerically calculating anisotropies using a fast Fourier transform (FFT). Eq (4.3) has also been derived in an elegant manner in ref [10].

## 5. Anisotropy Formulae for a Moving Point Mass

The simplest possible matter distribution is a single point mass. In Minkowski space a point mass produces no anisotropies if it is at rest with respect to the photon rest-frame, although it will contribute to the monopole anisotropy. More complicated anisotropies are produced by a moving point mass, (ref [9] §IV). We take the mass, $m$, to be moving ballistically with trajectory ${\bf x}' = {\bf x}_p(t)$, and velocity $\boldsymbol{\beta} = \dot{\bf x}_p(t)$. The stress-energy tensor is

$$\Theta_{\mu\nu} = m\gamma \begin{pmatrix} 1 & -\beta^i \\ -\beta^j & \beta^i\beta^j \end{pmatrix} \delta^{(3)}({\bf x} - {\bf x}_p(t)) \qquad (5.1)$$

where $\gamma = 1/\sqrt{1-\beta^2}$ is the Lorentz factor. For a collection of point masses not interacting with one another, the stress-energy tensor is a sum of terms such as in eq (5.1). Substituting this into eq (2.2) we find

$$\frac{\Delta T}{T}(\hat{\bf n}, {\bf x}_{\rm obs}, t_{\rm obs}) = m\gamma \left[ -2\frac{\boldsymbol{\beta}\cdot(X_{\rm obs}\hat{\bf n} + {\bf X}_{\rm obs})}{X_{\rm obs} + \hat{\bf n}\cdot{\bf X}_{\rm obs}} \frac{1 - \hat{\bf n}\cdot\boldsymbol{\beta}}{X_{\rm obs} + \boldsymbol{\beta}\cdot{\bf X}_{\rm obs}} + \frac{(1+\beta^2)(X_{\rm obs} - \hat{\bf n}\cdot{\bf X}_{\rm obs})}{X_{\rm obs}(X_{\rm obs} + \boldsymbol{\beta}\cdot{\bf X}_{\rm obs})} \right.$$
$$\left. + \frac{4\hat{\bf n}\cdot\boldsymbol{\beta}}{X_{\rm obs} + \boldsymbol{\beta}\cdot{\bf X}_{\rm obs}} - \frac{1+\beta^2}{\beta}\frac{\hat{\bf n}\cdot(\beta{\bf X}_{\rm obs} + X_{\rm obs}\boldsymbol{\beta})}{X_{\rm obs}(\beta X_{\rm obs} + \boldsymbol{\beta}\cdot{\bf X}_{\rm obs})} \right] \qquad (5.2)$$

for the full sky anisotropy pattern. An extra factor of $X_{\rm obs}/(X_{\rm obs} + \boldsymbol{\beta}\cdot{\bf X}_{\rm obs})$ comes from integrating over the $\delta$-function.

Using eq (3.3) we see that the monopole and dipole term components of eq (5.2) are given by

$$\Phi_{\rm eff} = -\frac{m\gamma}{X_{\rm obs} + \boldsymbol{\beta}\cdot{\bf X}_{\rm obs}} \left[ 1 + \beta^2 - 2\frac{\boldsymbol{\beta}\cdot{\bf X}_{\rm obs}}{X_{\rm obs}} + \frac{(\boldsymbol{\beta}\cdot{\bf X}_{\rm obs})^2}{X_{\rm obs}^2} \right]$$
$${\bf v}_{\rm eff} = -\frac{m\gamma}{X_{\rm obs} + \boldsymbol{\beta}\cdot{\bf X}_{\rm obs}} \left[ \frac{{\bf X}_{\rm obs}}{X_{\rm obs}}\left(1 + \beta^2 - 3\frac{\boldsymbol{\beta}\cdot{\bf X}_{\rm obs}}{X_{\rm obs}} + 2\frac{(\boldsymbol{\beta}\cdot{\bf X}_{\rm obs})^2}{X_{\rm obs}^2}\right) - \boldsymbol{\beta}\left(1 - \frac{\boldsymbol{\beta}\cdot{\bf X}_{\rm obs}}{X_{\rm obs}}\right) \right]. \qquad (5.3)$$
$$- m\gamma\frac{1+\beta^2}{\beta X_{\rm obs}}\frac{\beta{\bf X}_{\rm obs} + X_{\rm obs}\boldsymbol{\beta}}{\beta X_{\rm obs} + \boldsymbol{\beta}\cdot{\bf X}_{\rm obs}}$$

which can be compared to the Newtonian result

$$\Phi_{\text{Newt}} = -\frac{m}{X_{\text{obs}}}$$
$$\mathbf{v}_{\text{Newt}} = -\frac{m}{\beta}\frac{\beta\mathbf{X}_{\text{obs}} + X_{\text{obs}}\boldsymbol{\beta}}{X_{\text{obs}}(\beta X_{\text{obs}} + \boldsymbol{\beta}\cdot\mathbf{X}_{\text{obs}})} \quad (5.4)$$

which can be gotten by either taking the $\beta \ll 1$ limit of eq (5.4) or by doing a Newtonian analysis. Note that in keeping with action-at-a-distance the $X_{\text{obs}}$ in eq (5.4) is the distance to the point at the time of observation rather than on the past light-cone, however these two definitions of $\mathbf{X}_{\text{obs}}$ coincide for $\beta \ll 1$.

The small angle limit of the anisotropy pattern is given by eq (4.2), i.e.

$$\frac{\Delta T}{T}(\hat{\mathbf{n}}_\perp, \mathbf{x}_{\text{obs}}, t_{\text{obs}}) = -4m\gamma\frac{\hat{\mathbf{n}}_\perp\cdot\boldsymbol{\beta}}{\alpha} \quad (5.5)$$

which the result obtained in eq (4.1) of ref [9]. If the mass point is static, there is no anisotropy which is the same result as for the cosmological case (ref [8]).

## 6. Anisotropy Formulae for Cosmic Strings

Let us apply the formulae of the §2 to the case of cosmic strings (see ref [12] for a review). Cosmic strings are linear concentrations of mass density which in its rest frame has a tension and linear energy density which are both equal to the same constant, usually referred to as $\mu$. To describe the string we follow ref [9] and set up conformal coordinates, $\sigma$ and $t$ on the string world-sheet where $t$ is the usual time coordinate an $\sigma$ labels the position on the string. The string trajectory is then described by the function $\mathbf{r}(\sigma, t)$ and the equations of motion are

$$\dot{\mathbf{r}}\cdot\mathbf{r}' = 0 \qquad |\dot{\mathbf{r}}|^2 + |\mathbf{r}'|^2 = 1 \qquad \ddot{\mathbf{r}} - \mathbf{r}'' = 0 \quad (6.1)$$

where $\dot{}$ and $'$ refer to differentiation with respect to $t$ and $\sigma$, respectively. The stress-energy tensor is

$$\Theta_{\mu\nu}(x^\mu) = \oint d\sigma\, \tilde{\Theta}_{\mu\nu}(\sigma, t)\delta^{(3)}(\mathbf{x} - \mathbf{r}(\sigma, t)) \qquad \tilde{\Theta}_{\mu\nu}(\sigma, t) = \mu\begin{pmatrix} 1 & -\dot{r}^i \\ -\dot{r}^j & \dot{r}^i\dot{r}^j - r'^i r'^j \end{pmatrix}. \quad (6.2)$$

Since each $|\dot{\mathbf{r}}| < 1$, i.e. each string segment labeled by $\sigma$ moves at speed less than that of light, each segment will cross an observer's past light cone only once. Let us define the time of this crossing as $t_{\text{lc}}(\sigma)$ which is mathematically defined by the equation

$$t_{\text{obs}} - t_{\text{lc}}(\sigma) = |\mathbf{x}_{\text{obs}} - \mathbf{r}(\sigma, t_{\text{lc}}(\sigma))|. \quad (6.3)$$

Substituting eq (6.2) into eq (2.5) which drops pure monopole and dipole terms we obtain

$$\frac{\Delta T}{T}(\hat{\mathbf{n}}_0, \mathbf{x}_{\text{obs}}, t_{\text{obs}}) = -2\mu\oint d\sigma\frac{(X_{\text{obs}}\hat{\mathbf{n}}_0 + \mathbf{X}_{\text{obs}})\cdot[(1 + \hat{\mathbf{n}}_0\cdot\dot{\mathbf{r}})\dot{\mathbf{r}} - (\hat{\mathbf{n}}_0\cdot\mathbf{r}')\mathbf{r}']}{(X_{\text{obs}} + \hat{\mathbf{n}}_0\cdot\mathbf{X}_{\text{obs}})(X_{\text{obs}} + \dot{\mathbf{r}}\cdot\mathbf{X}_{\text{obs}})}\bigg|_{t=t_{\text{lc}}(\sigma)}. \quad (6.4)$$

The small-angle limit of this formulae is obtained by substituting eq (6.2) into eq (4.2) which yields

$$\frac{\Delta T}{T}(\hat{\mathbf{n}}_0, \mathbf{x}_{\text{obs}}) = -4\mu \oint d\sigma \frac{\mathbf{X}_{\text{obs}}^{\perp}}{X_{\text{obs}}^{\perp\,2}} \cdot \mathbf{u}(\sigma, t_{\text{lc}}(\sigma)) \qquad \mathbf{u} = \dot{\mathbf{r}} - \frac{(\hat{\mathbf{n}}_0 \cdot \mathbf{r}')\mathbf{r}'}{1 + \hat{\mathbf{n}}_0 \cdot \dot{\mathbf{r}}}. \qquad (6.5)$$

This is the same form as given in eq (6.16) of ref [9], but as we shall see the $\mathbf{u}$'s are different. The component of $\mathbf{u}$ parallel to $\hat{\mathbf{n}}_0$ does not matter in eq (6.5), and the remaining two components can be broken up into a piece parallel and a piece perpendicular to the projection of the string on the sky. Performing this decomposition with the help of formulae in §VI of ref [9] we find that we may rewrite eq (6.5) as

$$\frac{\Delta T}{T}(\hat{\mathbf{n}}_0, \mathbf{x}_{\text{obs}}) = 4\mu \oint \frac{d\mathbf{X}_{\text{obs}}^{\perp}}{X_{\text{obs}}^{\perp\,2}} \cdot \frac{(\hat{\mathbf{n}}_0 \cdot \mathbf{r}')\mathbf{X}_{\text{obs}}^{\perp} + (\hat{\mathbf{n}}_0 \cdot (\mathbf{r}' \times \dot{\mathbf{r}}))\hat{\mathbf{n}}_0 \times \mathbf{X}_{\text{obs}}^{\perp}}{|\mathbf{r}'|^2} \qquad (6.6)$$

where the $\hat{\mathbf{n}}_0 \cdot \mathbf{r}'$ term comes from the parallel component and the $\hat{\mathbf{n}}_0 \cdot (\mathbf{r}' \times \dot{\mathbf{r}})$ comes from the perpendicular component. The reason that this parallel-perpendicular decomposition is interesting is that the size of the perpendicular component gives the temperature discontinuity across the string while the parallel component does not contribute to the discontinuity at all. This can be understood in terms of the 2-d electrostatic analogy described in §6g of ref [9]. This analogy comes about since eq (6.5) is of the same form as the equation for the electric potential (given by $\Delta T/T$) for a "string" of electric dipoles linear dipole moment $\propto \mathbf{u}$. If we just had the component of $\mathbf{u}$ perpendicular to the projection then the string would act like a capacitor with all the positive charges on one side and the negative charges on the other. The dipole density would then give the jump in electric potential from one side of the capacitor to the other which represents the temperature discontinuity across a string. If the string were uniform then the parallel component of the $\mathbf{u}$ would not matter since each charge at each end of the dipole would be canceled by the opposite charge of the neighboring dipole on the string. However if the dipole density were not uniform then there would not be this exact cancellation and the parallel component would contribute to the anisotropy. Each excess charge then has a logarithmic potential profile in 2 dimensional potential theory. We shall see that this non-uniformity of the parallel component corresponds to curvature of the string. The importance of the parallel component was not fully explored in ref [9].

To demonstrate the importance of the parallel component let us examine it separately:

$$\frac{\Delta T^{\parallel}}{T}(\hat{\mathbf{n}}_0, \mathbf{x}_{\text{obs}}) = 4\mu \oint \frac{d\mathbf{X}_{\text{obs}}^{\perp}}{X_{\text{obs}}^{\perp\,2}} \cdot \frac{(\hat{\mathbf{n}}_0 \cdot \mathbf{r}')\mathbf{X}_{\text{obs}}^{\perp}}{|\mathbf{r}'|^2} = 4\mu \oint d(\ln X_{\text{obs}}^{\perp}) \frac{\hat{\mathbf{n}}_0 \cdot \mathbf{r}'}{|\mathbf{r}'|^2} = -4\mu \oint d\sigma \ln X_{\text{obs}}^{\perp} \frac{d}{d\sigma}\left[\frac{\hat{\mathbf{n}}_0 \cdot \mathbf{r}'}{|\mathbf{r}'|^2}\right]. \qquad (6.7)$$

It is easy to see for a straight string where $\dot{\mathbf{r}}$ and $\mathbf{r}'$ are constant along the string that this term contributes nothing. If $\mathbf{r}'$ is not constant then the string is either curved or will become curved since $|\dot{\mathbf{r}}|$ is not uniform along the string. This is also the condition for this term to give a non-zero contribution to the anisotropy. Since this term only depends on the modulus of $\mathbf{X}_{\text{obs}}^{\perp}$ it cannot contribute differently on one side of the string than on the other and therefore cannot contribute to the discontinuity across the string. The perpendicular component gives the temperature discontinuity and the analysis follow just as in ref [9].

**Comparison with Stebbins (1988)**

The anisotropy formula (6.5) is same in form to the one in eq (6.16) of ref [9] but instead of **u** ref [9] uses

$$\mathbf{u}^{S88} = \left(1 - \frac{(\hat{\mathbf{n}}_0 \cdot \mathbf{r}')^2}{(1+\hat{\mathbf{n}}_0 \cdot \dot{\mathbf{r}})^2}\right)\dot{\mathbf{r}}. \tag{6.8}$$

The difference between the two expressions is

$$\delta\frac{\Delta T}{T} \equiv \frac{\Delta T}{T}\bigg|_{SV93} - \frac{\Delta T}{T}\bigg|_{S88} = 4\mu \oint d\sigma \frac{\hat{\mathbf{n}}_0 \cdot \mathbf{r}'}{1+\hat{\mathbf{n}}_0 \cdot \dot{\mathbf{r}}} \frac{\mathbf{X}^\perp_{\text{obs}}}{X^{\perp\,2}_{\text{obs}}} \cdot \left(\mathbf{r}' - \frac{(\hat{\mathbf{n}}_0 \cdot \mathbf{r}')\dot{\mathbf{r}}}{1+\hat{\mathbf{n}}_0 \cdot \dot{\mathbf{r}}}\right) = -4\mu \oint d\sigma \ln X^\perp_{\text{obs}} \frac{d^2 t_{\text{lc}}(\sigma)}{d\sigma^2}. \tag{6.9}$$

The formulae are clearly different since $t_{\text{lc}}$ can have a non-zero 2nd-derivative. The expression derived in this section is correct and the formula of ref [9] is incorrect. A very simple derivation of the correct formula is given by Hindmarsh in ref [10].

If we decompose $\mathbf{u}^{S88}$ as in eq (6.6) we find that

$$\frac{\Delta T}{T}(\hat{\mathbf{n}}_0, \mathbf{x}_{\text{obs}}) = 4\mu \oint \frac{d\mathbf{X}^\perp_{\text{obs}}}{X^{\perp\,2}_{\text{obs}}} \cdot \left[\left(1 - \frac{|\mathbf{r}'|^2}{1+\hat{\mathbf{n}}_0 \cdot \dot{\mathbf{r}}}\right)\frac{\hat{\mathbf{n}}_0 \cdot \mathbf{r}'}{|\mathbf{r}'|^2}\mathbf{X}^\perp_{\text{obs}} + \frac{(\hat{\mathbf{n}}_0 \cdot (\mathbf{r}' \times \dot{\mathbf{r}}))}{|\mathbf{r}'|^2}\hat{\mathbf{n}}_0 \times \mathbf{X}^\perp_{\text{obs}}\right]. \tag{6.10}$$

We see that ref [9] obtained the correct formula for the perpendicular component but underestimated the parallel component by a factor

$$0 \leq 1 - \frac{|\mathbf{r}'|^2}{1+\hat{\mathbf{n}}_0 \cdot \dot{\mathbf{r}}} = \frac{|\dot{\mathbf{r}}|^2 + \hat{\mathbf{n}}_0 \cdot \dot{\mathbf{r}}}{1+\hat{\mathbf{n}}_0 \cdot \dot{\mathbf{r}}} \leq 1. \tag{6.11}$$

Since the perpendicular component was correct in ref [9] the result for the discontinuity across a moving string was also correct. However we see that anisotropy from the parallel component is underestimated by a velocity dependent factor. For ultra-relativistic velocities the factor is unity but for non-relativistic velocities the factor can be arbitrarily small. The large tension of strings cause them to have relativistic velocities, and we estimate that on average the parallel component is underestimated by about a factor of 2 in ref [9]. This does not mean that the total anisotropy obtained using eq (6.10) will be suppressed by this factor since the perpendicular component is correct. In fact it is clear that the small scale anisotropies are dominated by the discontinuity since the logarithmic term in eq (6.7) does not have much small scale power. Determining just the magnitude of the correction to the total anisotropy requires more study. We do not expect any of the qualitative results obtained to be effected by this correction.

**Horizons and Straight Strings**

The simplest string configuration is a static infinite straight string. It is well known that there exists a metric around such a string which is static and has an angular deficit around the string (see ref [13]). One would not expect any anisotropies in a static metric, yet if one were to substitute the stress-energy for such a string in eq (2.2) one would obtain a nonzero result. One cannot apply the our formalism to this case since eqs (2.3-4) are not satisfied and, in particular, the the spacetime is not asymptotically flat. There is a non-zero anisotropy produced by an infinite straight string which is created at an early time in an expanding universe as shown in ref [8].

As described in that paper, the cause of the anisotropies, is intimately related to the presence of horizons: the angular deficit is not present at large distances from the string and the information of the presence of the string will propagate outward at the speed of light. This time changing component of gravitational field does create anisotropies. To further illustrate that the cause of anisotropies is due to horizons we may try to mimic the cosmological situation in Minkowski space.

In empty Minkowski space there is no matter from which to make a string and in any case the making a gauge string would violate conservation of a topological charge of the gauge fields. Nevertheless, in the context of General Relativity on can create a string in Minkowski space if, in order to obey energy-momentum conservation, one remove energy from the vacuum. Following cosmological terminology we refer to the required energy deficit in the vacuum as "compensation". We will assume that the compensation takes the form of pressureless dust, and hence the string and the compensation remain superposed in the same place. Note that unlike normal dust the compensation has negative not positive energy density. The stress-energy tensor of this hypothetical configuration is

$$\Theta_{\mu\nu}(\mathbf{x},t) = \mu \begin{pmatrix} 0 & 0 & 0 & 0 \\ 0 & 0 & 0 & 0 \\ 0 & 0 & 0 & 0 \\ 0 & 0 & 0 & -1 \end{pmatrix} \delta(x)\,\delta(y)\,\mathcal{H}(t) \tag{6.12}$$

where we have placed the string along the $z$-axis and turned on the string at $t = 0$ ($\mathcal{H}$ is the Heaviside function). If one substitutes this into eq (2.2) we find the anisotropy pattern for observers inside the strings horizon is given by

$$\frac{\Delta T}{T}(\hat{\mathbf{n}}) = -4\mu \sin^2\theta \int_0^{\frac{\sqrt{1-u^2}}{u}} \frac{dw}{\sqrt{1+w^2}} \frac{\cos\theta \cos\phi \sqrt{1+w^2} - 1}{\cos^2\theta\, w^2 - 2\cos\theta \cos\phi \sqrt{1+w^2} + 1 + \cos^2\theta \cos^2\phi}$$
$$- 2\mu \ln\frac{1+\sqrt{1-u^2}}{u} - 2\mu \frac{\cos\theta \cos\phi}{u} \tan^{-1}\frac{\sqrt{1-u^2}}{u} \qquad u \equiv \frac{r_{\text{obs}}}{t_{\text{obs}}} \tag{6.13}$$

and the anisotropy is, of course, zero outside the horizon. Here $r_{\text{obs}}$ is the distance of the string from the observer and thus $u < 1$ inside the string's horizon and $u > 1$ outside. We can compare the temperature pattern given by eq (6.13) to that for a compensated string in a matter-dominated universe as derived in ref [8]. It is natural to equate the observer time, $t_{\text{obs}}$, and distance, $r_{\text{obs}}$, with the conformal time and comoving distance in the cosmological case. Thus the variable $u$ in eq (6.13) is to be equated with $u$ in ref [8]. In fig 1 we show the anisotropy pattern from eq (6.13) and that for a cosmological string both with $u = 0.5$. While the two patterns are not identical, qualitatively the two are very similar and we would argue that this is because the physics is essentially the same (for an explanation of some of the feature see ref [8]). This argues that the expansion of the universe plays no essential role in the anisotropy from a given seed configuration. The string configuration considered here is not moving, but as will shown in ref [4], the anisotropy from the motion in a cosmological setting is quite similar to that in Minkowski space.

## 7. Circular-Average Anisotropies and Spherically Symmetric Source Distributions

Even though formula for the anisotropy, eq (2.2), is much simpler than one might have expected one can not expect too many completely analytic expressions for anisotropy patterns. In most cases of interest the integrals will be too difficult and/or tedious to perform analytically. We can hope to make analytic progress with symmetrical matter configurations. In the rest of this paper we will examine anisotropies from spherically symmetric stress-energy configurations. We begin by averaging the anisotropy given by eq (2.2) over circles on the sky for arbitrary stress-energy configurations. Circular averages have interesting properties as was already exhibited in ref [9]. We then apply the circularly averaged formulae to a spherically symmetric stress-energy distribution.

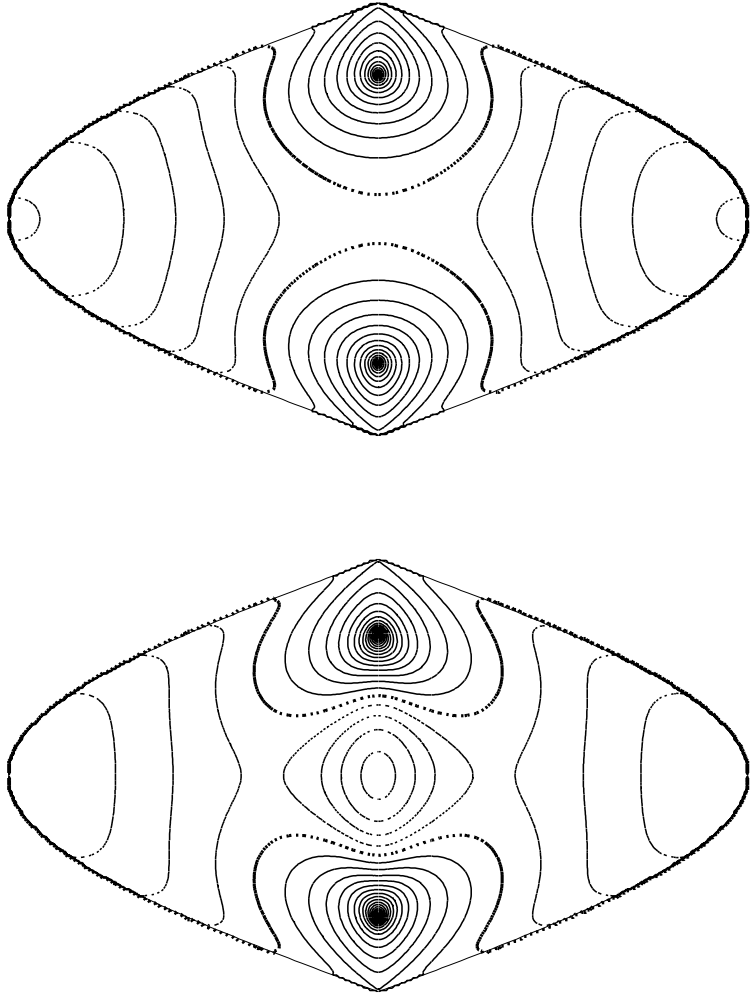

**FIGURE 1** Shown are two full sky contour maps of the anisotropy from two string configurations. The vertical coordinate gives the latitude while the horizontal coordinate give the longitude divided by the cosine of the latitude. This is an equal area projection. The solid lines give positive contours, and the dotted lines give negative contours, while the thicker line gives the zero contour. The contour interval is $0.25G\mu/c^2$ in the lower map and $G\mu/c^2$ in the upper map. In both maps the projection of the string lies along the straight line connecting the two hot-spots at the top and bottom of the maps. The lower pattern is from a string in a matter dominated universe, as calculated in ref [8]. The upper pattern is from a compensated string created at a finite time in Minkowski space (see text). Both configurations create a time changing gravitational field as the "information" of the string propagates outward. It is this effect of horizons which leads to the common features in both maps.

**Circular Averages of Temperature**

In eq (4.6) of ref [9] it was shown that in the small-angle approximation that the average temperature on a circle on the sky is independent of the stress-energy contained within the circle. Here "within" means the projection of the stress-energy on our celestial sphere is within the circle. If the circle is *outside* of the projection of the entire stress-energy distribution on the sky then the circular average is zero. This result applies for an isolated stress-energy concentrations in the limit of where this matter is at large distance. One is effectively assuming zero boundary conditions at infinity (i.e. large angles). For finite distance mass concentrations one can expect the monopole and dipole anisotropies given by eq (3.3) to be non-zero. These two terms will provide a non-zero boundary condition to the small-angle approximation. Thus we do not expect a zero circular average for circles containing the entire mass distribution. The extension of this small-angle result to the finite angle anisotropies considered here is the following: Consider the set of concentric circles on the sky which are large enough to contain the entire stress-energy distribution. The dependence of the average anisotropy with the angular radius of these circles, $\theta$, is $A + B \cos \theta$ where $A$ and $B$ are independent of $\theta$. This is just the angular dependence of a monopole and dipole. Thus one could retain the zero circular average result of the small-angle approximation if one allowed oneself to subtract off the appropriate monopole and dipole. In general one might have expected an arbitrary $\theta$-dependence, but we see that one needs only to measure the circular average on two circles and one has determined the entire $\theta$-dependence. We now demonstrate this result.

If one averages the temperature given by eq (2.2) on a circle on the sky one finds

$$\left\langle \frac{\Delta T}{T}(\hat{\mathbf{n}}) \right\rangle_\phi = A + B \cos \theta + \int_\theta^\pi d\alpha \left[ C(\alpha) + D(\alpha) \cos \theta \right] \quad (7.1)$$

where $\theta$ is the angle from the center of the circle,

$$A = 2\pi \int_0^\infty X_{\text{obs}} \, dX_{\text{obs}} \int_0^\pi d\alpha \left[ -2(1 - \cos \alpha) \bar{\Theta}_\perp - 2 \sin \alpha \, \bar{\Theta}_\parallel + 2(1 - \cos \alpha) \bar{\Theta}_{\perp\parallel} \right.$$

$$\left. + \frac{(1 - \cos \alpha)^2}{\sin \alpha} \bar{\Theta}_{\perp\perp} + \sin \alpha \, (\bar{\Theta}_{00} + \bar{\Theta}) \right]$$

$$B = 2\pi \int_0^\infty X_{\text{obs}} \, dX_{\text{obs}} \int_0^\pi d\alpha \left[ -\sin \alpha \, (\bar{\Theta} - 3\bar{\Theta}_{\parallel\parallel}) + 4(1 - \cos \alpha) \bar{\Theta}_{\perp\parallel} \right.$$

$$+ \frac{(1 - \cos \alpha)^2}{\sin \alpha} \bar{\Theta}_{\perp\perp} - 4 \sin \alpha \, \bar{\Theta}_\parallel$$

$$\left. - \sin \alpha \cos \alpha \left( (\bar{\Theta}_{00} + \bar{\Theta}) - \frac{1}{X_{\text{obs}}} \int_{-\infty}^{t_{\text{obs}}} dt' \, (\bar{\Theta}_{00} + \bar{\Theta}) \bigg|_{t_{\text{obs}} = t'} \right) \right]$$
$$(7.2)$$

$$C(\alpha) = -4\pi \int_0^\infty X_{\text{obs}} \, dX_{\text{obs}} \left[ 2 \cos \alpha \, \bar{\Theta}_\perp - 2 \sin \alpha \, \bar{\Theta}_\parallel + 2 \bar{\Theta}_{\perp\parallel} - 2 \frac{\cos \alpha}{\sin \alpha} \bar{\Theta}_{\perp\perp} \right]$$

$$D(\alpha) = -4\pi \int_0^\infty X_{\text{obs}} \, dX_{\text{obs}} \left[ -\sin \alpha \, (\bar{\Theta} - 3\bar{\Theta}_{\parallel\parallel}) - 4 \cos \alpha \, \bar{\Theta}_{\perp\parallel} + \frac{\cos \alpha + 1}{\sin \alpha} \bar{\Theta}_{\perp\perp} \right]$$
$$(7.3)$$

and
$$\begin{aligned}
\bar{\Theta}_{00}(X_{\text{obs}}, \alpha) &= \frac{1}{2\pi} \int_{-\pi}^{\pi} d\psi \, \Theta_{00}(\mathbf{x}', t_{\text{obs}} - X_{\text{obs}}) \\
\bar{\Theta}_{\parallel}(X_{\text{obs}}, \alpha) &= \frac{1}{2\pi} \int_{-\pi}^{\pi} d\psi \, \hat{z}^i \Theta_{i0}(\mathbf{x}', t_{\text{obs}} - X_{\text{obs}}) \\
\bar{\Theta}_{\perp}(X_{\text{obs}}, \alpha) &= \frac{1}{2\pi} \int_{-\pi}^{\pi} d\psi \, (\cos\psi \hat{x}^i + \sin\psi \hat{y}^i) \Theta_{i0}(\mathbf{x}', t_{\text{obs}} - X_{\text{obs}}) \\
\bar{\Theta}(X_{\text{obs}}, \alpha) &= \frac{1}{2\pi} \int_{-\pi}^{\pi} d\psi \, \Theta_{ii}(\mathbf{x}', t_{\text{obs}} - X_{\text{obs}}) \\
\bar{\Theta}_{\parallel\parallel}(X_{\text{obs}}, \alpha) &= \frac{1}{2\pi} \int_{-\pi}^{\pi} d\psi \, \hat{z}^i \hat{z}^j \Theta_{ij}(\mathbf{x}', t_{\text{obs}} - X_{\text{obs}}) \\
\bar{\Theta}_{\parallel\perp}(X_{\text{obs}}, \alpha) &= \frac{1}{2\pi} \int_{-\pi}^{\pi} d\psi \, \hat{z}^i (\cos\psi \, \hat{x}^j + \sin\psi \hat{y}^j) \Theta_{ij}(\mathbf{x}', t_{\text{obs}} - X_{\text{obs}}) \\
\bar{\Theta}_{\perp\perp}(X_{\text{obs}}, \alpha) &= \frac{1}{2\pi} \int_{-\pi}^{\pi} d\psi \, [\cos 2\psi \, (\hat{x}^i \hat{x}^i - \hat{y}^i \hat{y}^i) + \sin 2\psi \, (\hat{x}^i \hat{y}^j + \hat{y}^i \hat{x}^j)] \Theta_{ij}(\mathbf{x}', t_{\text{obs}} - X_{\text{obs}})
\end{aligned} \tag{7.4}$$

The $A$ and $B$ terms in eq (7.1) behave exactly as a monopole and a dipole component of the anisotropy. The $C$ and $D$ terms will, in general, also give some contribution to the $l = 0, 1$ anisotropy but it will contain higher moments as well. From eqs (7.1-4) we see that if there is a range of $\alpha$ where there is no matter along the observer's line-of-sight, i.e. $\Theta_{\mu\nu} = 0$, then for that range of $\alpha$ we see that $C(\alpha) = D(\alpha) = 0$, and thus in the same range for $\theta$: $\langle \Delta T/T \rangle = a + b\cos\theta$, where $a$ and $b$ are $\theta$-independent. This demonstrates the claim stated above, i.e that the circularly averaged anisotropy behaves like a monopole and dipole outside any stress-energy distribution. If there is no matter within some angular range $\theta \in [\theta_*, \pi]$ then $C$ and $D$ will also contribute zero within that range and the anisotropy will be $A + B\cos\theta$.

**Spherical Mass Distributions**

In this section we have so far been calculating circular average of the anisotropy but have not assumed the matter is distributed in a circularly symmetric way. One particular class of circularly symmetric stress-energy configurations are spherically symmetric ones. Here we will consider a general time-dependent spherically symmetric matter distribution. Let $r$ be the radius and $t$ the time, and $\hat{\mathbf{r}}$ the outward pointing radial unit vector at each point. The most general stress-energy tensor is

$$\Theta_{00} = \rho(r,t) \qquad \Theta_{i0} = -V(r,t) \hat{r}^i \qquad \Theta_{ij} = p(r,t)\delta_{ij} + \Pi(r,t)(3\hat{r}^i \hat{r}^j - \delta_{ij}) \tag{7.5}$$

where $\rho$ is the density, $S$ the radial momentum flux, $p$ the isotropic pressure, and $\Pi$ gives the anisotropic component to the pressure tensor. The only constraints on these 4 functions is that they obey energy and momentum conservation. The energy and momentum conservation laws reads

$$\dot{\rho} + \frac{\partial}{\partial r}V + \frac{2}{r}V = 0 \qquad \dot{V} + \frac{\partial}{\partial r}(2\Pi + p) + \frac{6}{r}\Pi = 0. \tag{7.6}$$

The total mass of this matter distribution is

$$M = 4\pi \int_0^\infty \rho(r,t) \, r^2 \, dr \tag{7.7}$$

which is constant in time.

This spherically symmetric matter distribution will induce a circularly symmetric anisotropy pattern on the sky. In this case one can average around the axis of symmetry and not lose any information since there is no azimuthal dependence. Thus the $\theta$-dependence given by the formulae in the previous subsection gives the full temperature pattern. Let $r_{\text{obs}}$ be the distance of the observer from the center of symmetry of the matter distribution in which case eq (7.4) becomes

$$\begin{aligned}
\bar{\Theta}_{00} &= \rho(r, t_{\text{obs}} - X_{\text{obs}}) \\
\bar{\Theta}_{\parallel} &= \frac{r_{\text{obs}} - X_{\text{obs}} \cos \alpha}{r} V(r, t_{\text{obs}} - X_{\text{obs}}) \\
\bar{\Theta}_{\perp} &= -\frac{X_{\text{obs}} \sin \alpha}{r} V(r, t_{\text{obs}} - X_{\text{obs}}) \\
\bar{\Theta} &= 3p(r, t_{\text{obs}} - X_{\text{obs}}) \\
\bar{\Theta}_{\parallel\parallel} &= p(r, t_{\text{obs}} - X_{\text{obs}}) + 3\frac{(r_{\text{obs}} - X_{\text{obs}} \cos \alpha)^2}{r^2} \Pi(r, t_{\text{obs}} - X_{\text{obs}}) \\
\bar{\Theta}_{\parallel\perp} &= -3\frac{(r_{\text{obs}} - X_{\text{obs}} \cos \alpha) X_{\text{obs}} \sin \alpha}{r^2} \Pi(r, t_{\text{obs}} - X_{\text{obs}}) \\
\bar{\Theta}_{\perp\perp} &= 3\frac{X_{\text{obs}}^2 \sin^2 \alpha}{r^2} \Pi(r, t_{\text{obs}} - X_{\text{obs}})
\end{aligned} \tag{7.8}$$

$$r \equiv \sqrt{X_{\text{obs}}^2 + r_{\text{obs}}^2 - 2 X_{\text{obs}} r_{\text{obs}} \cos \alpha}$$

where the distance, $r$, from the center of symmetry to a field point, $\mathbf{x}'$. Thus $C$ and $D$ are given by

$$\begin{aligned}
C(\alpha) &= 8\pi \sin \alpha \int_0^\infty X_{\text{obs}} \, dX_{\text{obs}} \frac{r_{\text{obs}}}{r} \left[ V(r, t_{\text{obs}} - X_{\text{obs}}) + 3 \frac{X_{\text{obs}}}{r} \Pi(r, t_{\text{obs}} - X_{\text{obs}}) \right] \\
D(\alpha) &= -12\pi \int_0^\infty X_{\text{obs}} \, dX_{\text{obs}} \left[ 1 + \frac{2 r_{\text{obs}}^2 + X_{\text{obs}}^2 (1 - \cos \alpha) \cos \alpha}{r^2} \right] \Pi(r, t_{\text{obs}} - X_{\text{obs}}) \sin \alpha
\end{aligned} \tag{7.9}$$

We will use these expressions to calculate the anistropy for a few specific stress-energy configurations.

## 8. Anisotropy from a Collapsing Texture Knot

Now we calculate the anisotropy around an unwinding knot of $N = 4$ cosmic texture (ref [14]). There is some debate over exactly what volume of space contains knots which unwind (ref [15]), and as a result, over whether the pattern of anisotropy is dominated by the fully unwinding knots, or by partially unwinding knots and the field gradients which cover all space (ref [16]). If a fully unwinding knot were relatively close to the observer, this being the limit in which our formalism is applicable, one expects the knot to dominate the anisotropy around this observer, and we show below the expected pattern.

Let $\mathcal{V}$ be the vacuum expectation value of the scalar field which gives rise to texture, and consider the spherically symmetric self-similar solution found in the nonlinear sigma model approximation to the actual texture field ref [3,17]. If the knot collapses at time $t_c$, the spherically

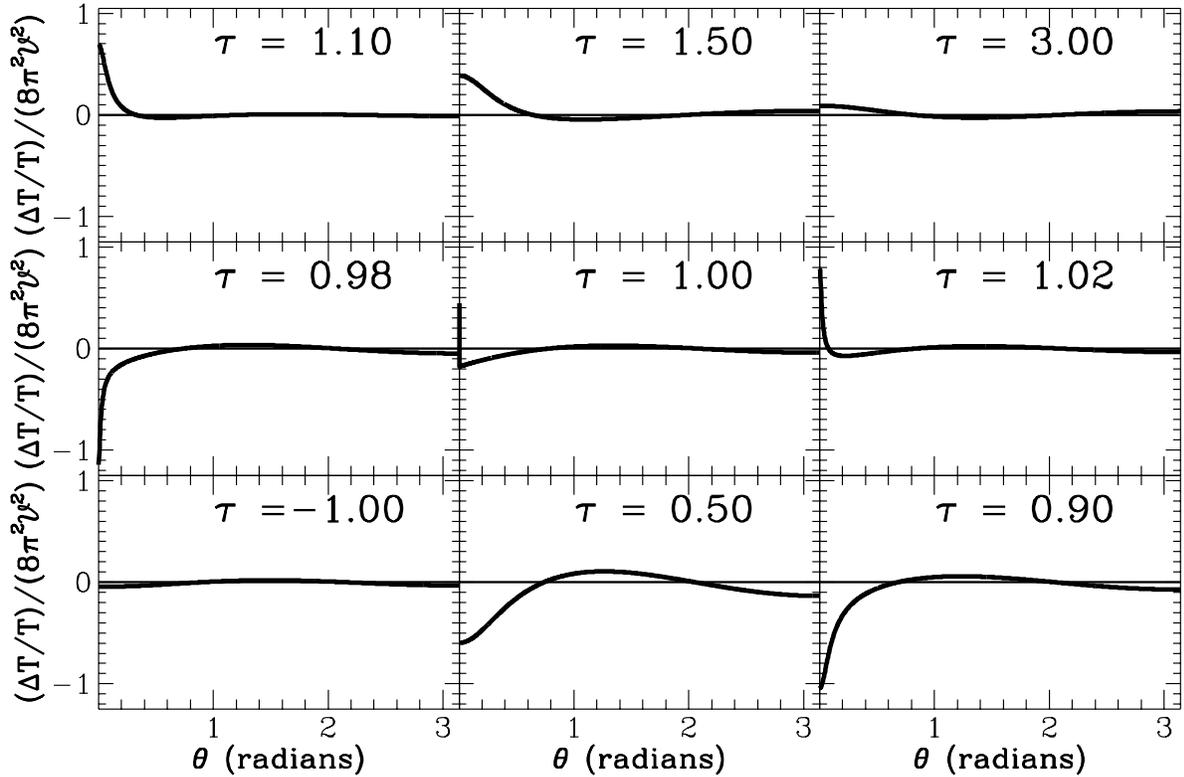

**FIGURE 2** Shown is the anisotropy from a collapsing texture knot as a function of the angle from the directions toward the center of the knot for various times before and after collapse. The angle $\theta$ and time $\tau$ are as in eq (8.2) but here the monopole and dipole component of the anisotropy is has been subtracted off. The observer sees the collapse at $\tau = 1$. In the small-angle approximation the pattern would be anti-symmetric about $\tau = 1$ and the curves would approach $-1$ for $\tau < 1$ and $+1$ for $\tau > 1$.

symmetric stress-energy tensor is,

$$\Theta_{00} = 2\mathcal{V}^2 \frac{r^2 + 3(t - t_c)^2}{(r^2 + (t - t_c)^2)^2} \qquad \Theta_{0i} = -4\mathcal{V}^2 \frac{r(t - t_c)}{(r^2 + (t - t_c)^2)^2} \hat{\mathbf{r}} \qquad \Theta_{ij} = 2\mathcal{V}^2 \frac{r^2 - (t - t_c)^2}{(r^2 + (t - t_c)^2)^2} \delta_{ij},$$
(8.1)

where $r$ is the distance from the centre of the texture knot. Let $d$ be the distance of the observer from the texture center. The self-similar nature of the knot distribution shows up in that the anisotropy is a function only of the ratio $\tau = (t_{\text{obs}} - t_c)/d$ and the angle from the texture center, $\theta$. Here $\tau$ will label the time of observation with $\tau = 1$ being the time at which the observer "sees" the texture collapse. The spherical symmetry allows us to use the specialized result of eq (7.9) to compute the anisotropy, which is

$$\frac{\Delta T}{T}(\hat{\mathbf{n}}, \mathbf{x}_{\text{obs}}, t_{\text{obs}}) = 8\pi \mathcal{V}^2 \left[ \frac{\tau - \cos\theta}{\sqrt{(\tau - \cos\theta)^2 + 2\sin^2\theta}} \left( \frac{\pi}{2} + \sin^{-1} \frac{\tau + \cos\theta}{\sqrt{2(\tau^2 + 1)}} \right) \right.$$

$$\left. - \text{sgn}(\tau + 1) \left( \frac{\pi}{2} + \sin^{-1} \frac{\tau - 1}{\sqrt{2(\tau^2 + 1)}} \right) \right]$$
(8.2)

where $\text{sgn}(x) = |x|/x$, and we have dropped the monopole and dipole terms and we have added the last term in square brackets, which contributes no angular dependence, to guarantee the boundary conditions of $\Delta T/T = 0$ at in the direction opposite to the collapsing knot. Fig 2 shows the anisotropy as a function of $\theta$ for various values of $\tau$, spanning times both before and after the collapse of the knot. The monopole and dipole components of the anisotropy have been subtracted. The small angle limit near the direction of texture collapse is formally

$$\frac{\Delta T}{T}(\hat{\mathbf{n}}, \mathbf{x}_{\text{obs}}, t_{\text{obs}}) \rightarrow 8\pi \mathcal{V}^2 \frac{\tau - \cos\theta}{\sqrt{2\sin^2\theta + (\tau - \cos\theta)^2}} \qquad (8.3)$$

where we have again adjusted the zero-point so that the anisotropy goes to zero at large angles from the collapsing knot. One could also obtained eq (8.3) by substituting eq (8.1) into eq (4.2) and this formula is the same as obtained in ref [3].

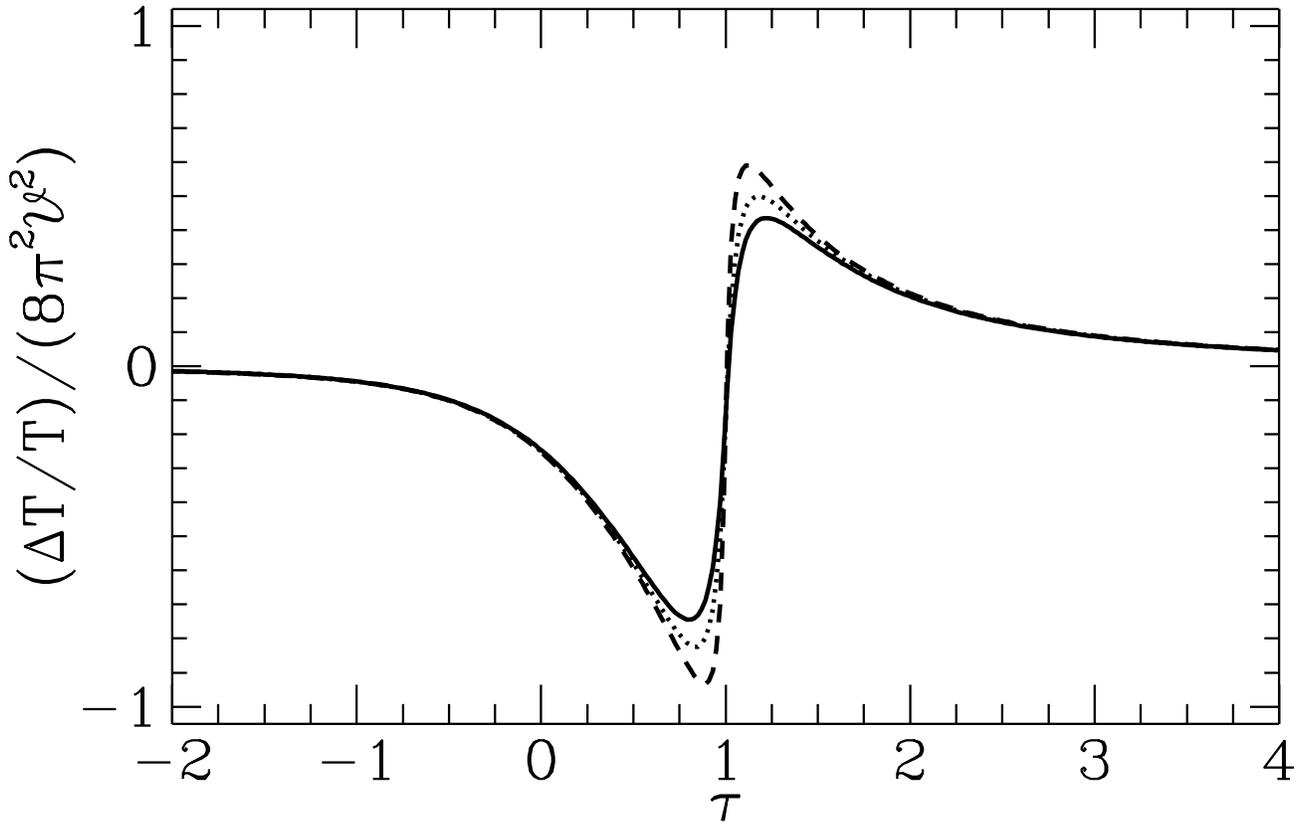

**FIGURE 3** Shown is the smoothed anisotropy from a collapsing texture knot as a function of time. In particular we plot $\Delta T/T(\theta = 0, \tau)$ by convolving the temperature pattern of eq (8.2) with a Gaussian beam and subtracting the monopole and dipole. The smoothing lengths are given by FWHM=10° (*solid curve*), 7° (*dotted curve*), and 3.8° (*dashed curve*) corresponding to smoothed DMR, unsmoothed DMR, and FIRS, respectively (see ref [1]).

Both the large-angle and small-angle formulae give a magnitude of the temperature jump for photons passing through the texture center just before and just after $\tau = 1$ of $16\pi^2 \mathcal{V}^2$. While eqs (8.2&3) agree in the appropriate limit, there are some qualitative differences between them. In particular we see that the pattern in the small-angle approximation is anti-symmetric about the $\tau = 1$ while there is no such symmetry in the large-angle formula. For example the small-angle formulae has $\Delta T/T(\theta = 0)$ go from $-8\pi^2 \mathcal{V}^2$ to $+8\pi^2 \mathcal{V}^2$ at $\tau = 1$ while the large-angle formulae has $\Delta T/T(\theta = 0)$ go from $-12\pi^2 \mathcal{V}^2$ to $+4\pi^2 \mathcal{V}^2$. One could reduce the asymmetry greatly by appropriate choice of the dipole however one cannot get rid of it completely as illustrated in fig 2. Here we plot monopole- and dipole-subtracted $\Delta T/T$ of eq (8.2) as a function of $\theta$ for a variety of values $\tau$. The values of $\tau$ shown are symmetrically placed about $\tau = 1$ and we see that there remains an asymmetry.

In fig 3 we have taken the monopole- and dipole-subtracted patterns, convolved them with Gaussian beams, and evaluated the anisotropy at the center of the texture spot for a range of times during the texture collapse. We see that the asymmetry mentioned above causes the magnitude of the coldest spot anisotropy to be larger than the magnitude of hottest spot anisotropy. Furthermore we find that the anisotropies drop rapidly from their extremal values. These corrections to the small-angle approximation leads to smaller expectations for the amplitudes of texture spots. The maximal amplitude at small angles, $\pm 8\pi^2 \mathcal{V}^2$, are not obtained even for the spherical self-similar texture solution. More realistic configurations are likely to lead to even smaller anisotropies (ref [16]).

## 9. Summary

In this paper we have presented some of the phenomenology of light passing through a time-changing gravitational field. While the basic equations of the passage of light through arbitrary gravitational fields have been known for nearly half a century, most of the interesting phenomenology involves applications to nearly Newtonian systems such as the Solar System, where it is necessary in the understanding of pulsar timing. In this paper we have given the general solution for the energy (or frequency) shift of massless particles passing through the gravitational field of an arbitrary mass distribution in the weak field limit. In §2 we have expressed the result as the convolution of the stress-energy of the mass distribution with certain Green functions. Since the stress-energy tensor is constrained to obey energy-momentum conservation, there are a variety of different but equally valid sets of Green functions. One interesting property of the Green functions given here is that the pattern of energy shifts of photons which arrive at a given point in space-time depends essentially only on the stress-energy distribution on the past light-cone of that point. This in spite of the fact that an infinite variety of different stress-energy histories would obtain those values on this light-cone. This is an extension of the similar result found in the small-angle approximation (see refs [9,10]).

In §3 we further elucidate the properties of the Green functions. We show the dependence on the two different direction angles must be expressible as the sum of a *finite*, 5-term Fourier series of one of the angles. This is a consequence of the spin-2 nature of gravitational fields and must apply in any isotropic space such as a Friedmann-Robertson-Walker (FRW) cosmology. We also extract the monopole and dipole components of the angular pattern to give the effective Newtonian potential and acceleration. In §4 we give the small-angle limit of our formula which which concurs with the result presented in ref [10]. Various properties of this form of the result can be expressed in terms of 2-dimensional potential theory as explained in ref [9]. In §5 we apply our Green function to a moving point mass and in §6 to cosmic strings. We correct an error in ref [9] which leads to (possibly small) underestimation of the anisotropy from strings. We also compare the full-sky pattern from a "compensated" string in Minkowski space to the similar configuration in an expanding universe. We argue that the qualitative similarities is due to the presence of horizons in both configurations.

In §7 we calculate the circular average of the anisotropy pattern on the sky. In ref [9] it was shown, using the small-angle approximation, that outside of the projected matter distribution that such averages are zero. Similar properties are also found outside of the small-angle approximation. The equation for the circularly averaged anisotropy may be used to simplify the calculation of the anisotropy pattern from spherical matter distributions. In §8 we apply our results to the collapsing texture knot solution of ref [3]. We show that the time symmetry in the pattern in the small angle approximation breaks down at large angles leading to brighter cold-spots than hot-spots. The amplitudes of the spots were also found to be decreased with respect to the small-angle approximation.

The Green functions found here are for matter distributions which yield small perturbations about Minkowski space and one cannot apply them to a general cosmological setting. In a separate paper (ref [4]) we will give the Green functions for small perturbations about a flat matter-dominated FRW cosmology. The Green functions are significantly more complicated in that case, yet they retain many of the qualitative features of the Minkowski Green functions given here. Furthermore many of the peculiar features of anisotropy patterns from seeds, such as the discontinuity of cosmic strings or the hot- and cold-spots produced by cosmic textures can be demonstrated in a Minkowski setting.

**Acknowledgements** AS was supported in part by the DOE and NASA grant NAGW-2381 at Fermilab. SVR was supported by the National Research Council.

## Appendix

To calculate the the anisotropy pattern for the congruence of freely-falling observers defined above it is easiest to use synchronous coordinates which comove with the observers. In these coordinate each observer follows a trajectory of fixed spatial coordinates, $\mathbf{x}$, while the temporal coordinate, $t$, gives the proper time of experienced by the observer. The metric of such a coordinate system has $h_{00} = h_{i0} = 0$. The fractional energy shift of eq (2.1) is given by Sachs-Wolfe integral which is the temporal component of the linearized geodesic equation for the photon. In comoving synchronous coordinates the Sachs-Wolfe integral reads

$$\frac{\Delta T}{T}(\hat{\mathbf{n}}, \mathbf{x}_{\rm obs}, t_{\rm obs}) = -\frac{1}{2}\int_{-\infty}^{t_{\rm obs}} dt\, \hat{n}^i \hat{n}^j \dot{h}_{ij}(\mathbf{x}_\gamma(t), t) \qquad \mathbf{x}_\gamma(t) = (t_{\rm obs} - t)\hat{\mathbf{n}} + \mathbf{x}_{\rm obs}. \qquad (\text{A1})$$

Since we are interested in the anisotropy to 1st order in $h_{ij}$ we have taken the photon trajectory as in flat space, i.e. to zeroth order in $h_{ij}$. In the limit of non-relativistic sources this equation is equivalent to the more familiar Newtonian eq (1.1). Since Minkowski space is conformally related to a flat FRW cosmology eq (A1) is equally valid in such a cosmology if one takes $t$ to be the conformal time and $\mathbf{x}$ the comoving coordinates. However the solutions we present below do not carry over to the cosmological case since the solutions for $h_{ij}$ are not the same.

Equation (A1) tells us that the temperature shift suffered by a photon traveling in any direction is given by the time derivative of the metric perturbations along its trajectory. One may solve the linearized Einstein's equation to obtain the metric perturbation as an integral over the time history of the source stress-energy, say by taking the Minkowski limit of the cosmological eqs in §5 of ref [8], obtaining

$$h_{ij}(\mathbf{x},t) = 4\int_{-\infty}^{t} dt' \int d^3x' \frac{\delta(t-t'-X)}{X} \Theta_{\{ij\}}(\mathbf{x}',t')$$

$$-8\int_{-\infty}^{t} dt' \int d^3x' \int_{-\infty}^{t'} dt'' \frac{\delta(t-t'-X)}{X} \Theta_{o\{j,i\}}(\mathbf{x}',t'')$$

$$+2\int_{-\infty}^{t} dt' \int d^3x' \frac{\delta(t-t'-X)}{X} \int_{-\infty}^{t'} dt''(t'-t'')\Theta_{+,\{ij\}}(\mathbf{x}',t'') \quad (A2)$$

$$+\frac{8\pi}{3} \int_{-\infty}^{t} dt'(t-t')\Theta_{+}(\mathbf{x}',t')$$

where $X = |\mathbf{x}-\mathbf{x}'|$, $\Theta_{+} = \Theta_{00} + \Theta_{ii}$, and $\{ij\}$ gives the symmetric traceless part, i.e.

$$f_{\{ij\}} = \frac{1}{2}\left(f_{ij} + f_{ji} - \frac{2}{3}\delta_{ij}f_{kk}\right). \quad (A3)$$

After substituting eq (A3) into eq (A2) one obtains our general solution, eq(2.2), using the the assumptions of eqs (2.3-4) and the equations of energy-momentum conservation:

$$\dot{\Theta}_{00} - \Theta_{0i,i} = 0 \qquad \dot{\Theta}_{0i} - \Theta_{ij,j} = 0. \quad (A4)$$

On the whole the derivation is straightforward, however at one point we use eq (A4) to derive an identity which allows us to integrate out a total divergence which greatly simplifies the result. Just such a procedure is used in refs [9,10,11] and for completeness we give the identity here:

$$\frac{d}{dt}\frac{(X_\gamma \hat{n}^i + X_\gamma^i)\Theta_{\alpha i}(\mathbf{x}',t-X_\gamma)}{X_\gamma(X_\gamma + \hat{\mathbf{n}}\cdot\mathbf{X}_\gamma)} = \frac{(X_\gamma \hat{n}^i + X_\gamma^i)\dot{\Theta}_{\alpha i}(\mathbf{x}',t-X_\gamma)}{X_\gamma^2} + \frac{X_\gamma^i \Theta_{\alpha i}(\mathbf{x}',t-X_\gamma)}{X_\gamma^3} \quad (A5)$$

where $\mathbf{X}_\gamma = \mathbf{x}_\gamma(t) - \mathbf{x}'$ and $X_\gamma = |\mathbf{X}_\gamma|$.